\documentclass{elsart}
\usepackage{epsfig}

\def\be{\begin{equation}}
\def\ee{\end{equation}}
\def\bea{\begin{eqnarray}}
\def\eea{\end{eqnarray}}
\def\beas{\begin{eqnarray*}}
\def\eeas{\end{eqnarray*}}
\def\beq{\begin{equation}}
\def\eeq{\end{equation}}
\def\bq{\begin{quote}}
\def\eq{\end{quote}}
\def\gappeq{\mathrel{\rlap {\raise.5ex\hbox{$>$}} {\lower.5ex\hbox{$\sim$}}}}
\def\lappeq{\mathrel{\rlap{\raise.5ex\hbox{$<$}} {\lower.5ex\hbox{$\sim$}}}}
\def\PR{{\it Phys.~Rev.~}}
\def\PRL{{\it Phys.~Rev.~Lett.~}}
\def\NP{{\it Nucl.~Phys.~}}

\def\PL{{\it Phys.~Lett.~}}

\def\ZP{{\it Zeit.~Phys.~}}
\def\JP{{\it Jour.~Phys.~}}

\def\EPJ{{\it Eur.~Phys.~J.~}}
\def\vol#1{{\bf #1}}
\def\vyp#1#2#3{\vol{#1} (#2) #3}

\def\epm#1#2{\hbox{${\lower1pt\hbox{$\scriptstyle +#1$}}
\atop {\raise1pt\hbox{$\scriptstyle -#2$}}$}}

\def\gsim{\mathrel{\rlap{\lower4pt\hbox{\hskip1pt$\sim$}}
    \raise1pt\hbox{$>$}}}         

\def\etal{{\it et al.}}
\def\rhs{right hand side}

\def\frac#1#2{{{#1}\over {#2}}}
\def\half{\hbox{${1\over 2}$}}

\def\smallfrac#1#2{\hbox{${{#1}\over {#2}}$}}

\def\bq{\bar{q}}
\catcode`@=11 
\def\slash#1{\mathord{\mathpalette\c@ncel#1}}
 \def\c@ncel#1#2{\ooalign{$\hfil#1\mkern1mu/\hfil$\crcr$#1#2$}}
\def\lsim{\mathrel{\mathpalette\@versim<}}
\def\gsim{\mathrel{\mathpalette\@versim>}}
 \def\@versim#1#2{\lower0.2ex\vbox{\baselineskip\z@skip\lineskip\z@skip
       \lineskiplimit\z@\ialign{$\m@th#1\hfil##$\crcr#2\crcr\sim\crcr}}}
\catcode`@=12 

\begin{document}

\begin{frontmatter}
\title{ Flavor Decomposition of Nucleon Structure at a 
Neutrino Factory}
\author{Richard D. Ball\thanksref{label2}}
\thanks[label2]{Royal Society University Research Fellow}
\address{Department of Physics and Astronomy, 
University of Edinburgh,\\
Mayfield Road, Edinburgh EH9 3JZ, Scotland}
\author{Deborah A. Harris}
\address{Fermi National Accelerator Laboratory, Batavia, Illinois 60510, USA}
\author{Kevin S. McFarland} 
\address{Department of Physics and Astronomy, University of 
Rochester, \\ Rochester, New York 14627, USA} 

\begin{abstract}
We explore the possibilities for measuring the 
quark content of the proton and neutron using neutrino beams produced at a 
muon storage ring.  Because of the nature of the beams, 
small nuclear targets such as hydrogen and deuterium can be considered, 
as well as polarized targets.  The statistics expected from these  
targets are calculated using nominal 
muon storage ring intensities, and the resulting statistical 
errors on the numerous structure functions available are given, 
for both polarized and unpolarized targets.  
It is shown that with a relatively small target, the 
structure functions $F_2$, $xF_3$, $xg_1$ and $xg_5$ for 
neutrinos and antineutrinos
on protons and deuterium, either unpolarized or polarized, could 
be determined with excellent precision over most of the  
accessible kinematic range.
\end{abstract}
\end{frontmatter}

\section{Introduction}

Deep Inelastic Scattering (DIS) has long been the definitive process for 
the determination of the quark content of protons and neutrons.  
Charged lepton scattering has dominated the field in terms of 
precision determinations of the sum of the quark and antiquark 
distributions, and the associated gluon distribution,
but neutrino scattering has thus far contributed complementary 
measurements of the valence quark distributions, as well as 
measurements of the strange sea.  However, neutrino scattering  
has in the past been plagued by tiny interaction rates and large beam 
spot sizes, requiring targets on the order of several meters wide 
and several hundred tons to get appreciable statistics.  

\begin{figure}[t!] 	\hskip1cm                
\epsfig{file=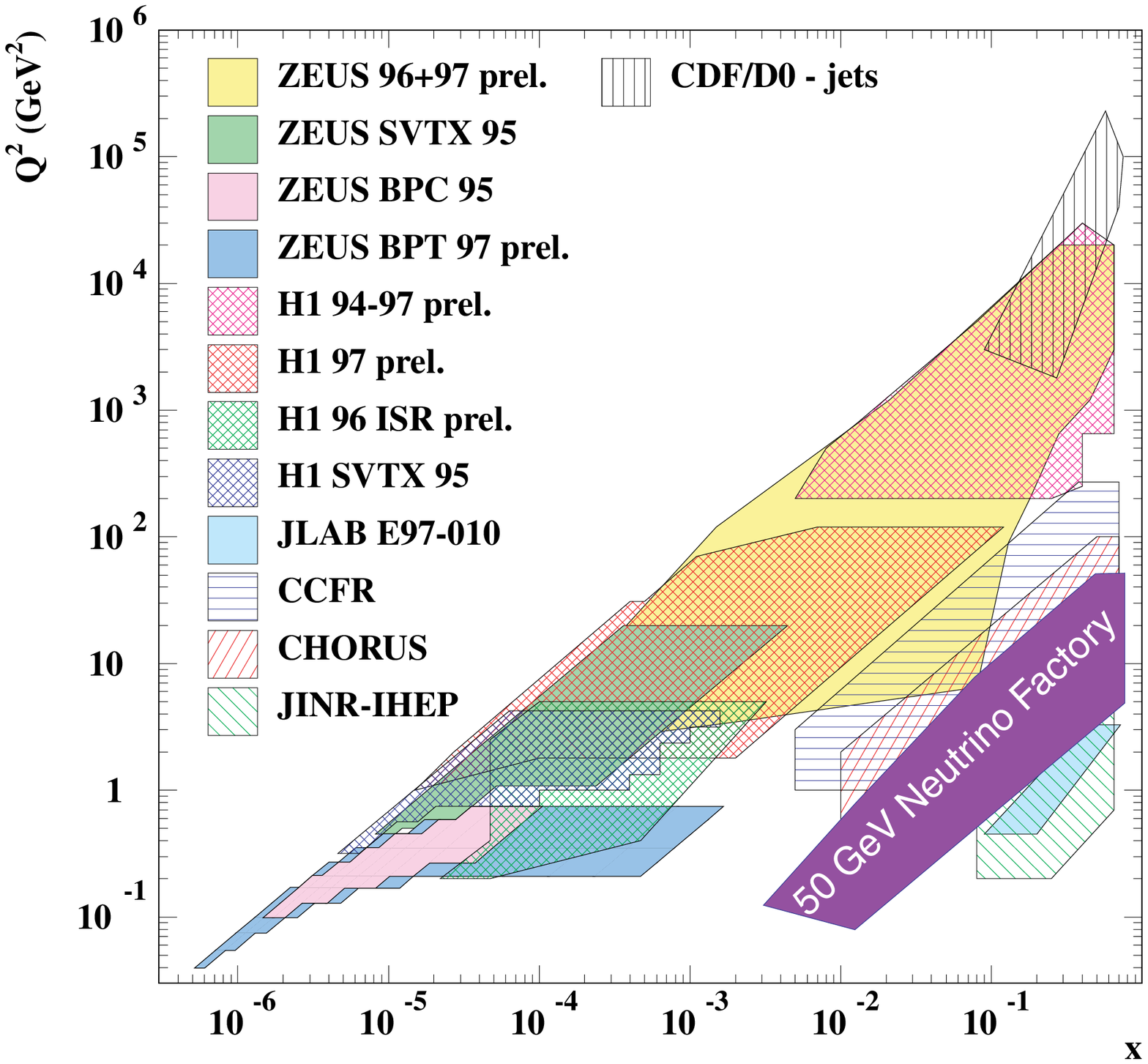,width=10cm,
bbllx=0pt,bblly=150pt,bburx=500pt,bbury=650pt}
\caption[]{\label{fig:kin}
The kinematic region in the $(x,Q^2)$ plane available at a $50$~GeV 
neutrino factory.}
\end{figure}

With the advent of a muon storage ring the flux of neutrinos at a near 
detector would be several orders of magnitude higher than at present 
experiments, and concentrated in a much smaller spot size.  
Because of this one can now consider using 
compact hydrogen and deuterium targets, rather 
than iron. These targets have the advantage of allowing 
measurements of the valence 
quark distributions without nuclear effects, or conversely one can 
finally measure nuclear effects in valence 
quark distributions by comparing results using different targets.  
Many of these ideas (as well as other high-rate neutrino 
experiments at muon storage rings) 
are considered in references \cite{nufactexp},\cite{bkbook}, 
and \cite{fnalstudy}.  
Because it is expected that the storage ring will run in roughly 
equal running times in $\mu^+$ and $\mu^-$ mode, the fluxes  
for $\nu_e$ and $\bar\nu_e$ will be approximately equal, 
as will the fluxes 
for  $\bar\nu_\mu$ and $\nu_\mu$.  
In conventional beam neutrino experiments the dominant statistical 
error has been the antineutrino event rate, because the typical total 
antineutrino event rate (on the targets used) 
has been only 20-25\% of the neutrino event rate.  

Aside from just statistical considerations, the neutrino beams from a 
muon storage ring offer ways to lower previously important systematic
uncertainties as well.   
Because of the well-known incoming $\nu$ spectrum one has an accurate 
determination of the beam energy, as well as a calibration tool for the 
detector.  One also has an extremely pure neutrino beam in terms of 
sign selection; if $\mu^-(\mu^+)$ are circulating in the ring the fluxes
available are $\nu_\mu$ and $\bar\nu_e$ ($\bar\nu_\mu$ and $\nu_e$ ).  
By identifying the flavor of the final state lepton in a 
charged current interaction one knows if the initial lepton was a 
neutrino or an antineutrino.  Finally, again because of the small 
beam spot size, polarized $\nu$-DIS experiments could be  
performed for the first time, allowing a full flavor decomposition of the 
nucleon spin.  Because these measurements would still be inclusive 
DIS measurements, they would not suffer from the fragmentation 
uncertainties intrinsic to semi-inclusive measurements such as 
those made at HERMES.  

One disadvantage of the neutrino beams from a muon storage ring is that 
because the rings are expected to operate at a relatively low beam 
energy ($30$ to $50$ GeV), a 
lower and smaller range of momentum transfers will be available.  
The expected kinematic range for a $50$~GeV beam is shown in 
fig\ref{fig:kin}. 

The remainder of this paper consists of a description of the theoretical 
framework by which the quark structure of the nucleon can be measured, 
a brief discussion of what the neutrino target and detector would  
look like, and finally, preliminary leading order estimates of 
expected uncertainties for a variety of stucture functions.  

\section{Unpolarized Structure Functions}

Unpolarized charged current structure functions are defined 
through the decomposition of unpolarized differential charged current 
cross-sections into invariant functions of 
the momentum of the struck quark ($x$) and and the momentum transfer
squared of the $W$ boson ($Q^2$): the standard definitions give
\beq
\smallfrac{d^2\sigma^{\nu\bar\nu}}{dx dy} = 
\smallfrac{G_F^2 S}{2\pi(1+Q^2/M_W^2)^2}
[(1-y)F_2^{\nu,\bar\nu} 
+ y^2 xF_1^{\nu,\bar\nu} 
\pm y(1-\smallfrac{y}{2})xF_3^{\nu,\bar\nu}],\label{sfn}
\eeq
where $S=2mE$ is the centre-of-mass energy, $E$ is the neutrino 
beam energy, assumed to be $\gg m$, and the $\pm$ signs 
refer to the sign of the charged current: $W^+$ 
exchange for $\nu$ scattering and $W^-$ for $\bar\nu$. $y$ is
the fractional lepton energy loss, or $(E_\nu-E_\ell)/E_\nu$.  
In neutrino scattering, $x,y$, and $Q^2$ can all be determined 
simply by measuring the outgoing lepton energy and direction, 
and the hadronic energy in the event.  
There are then in principle six independent structure functions to be 
measured for every target. The  
proton structure functions are the same as those measured at higher $Q^2$ (see fig.\ref{fig:kin} in 
charged current $e^\pm p$ scattering at HERA.

In the parton model four of the these structure functions 
are related through the Callan-Gross relations $F_2=2xF_1$:
the longitudinal structure function $F_L=F_2-2xF_1$ begins at 
$O(\alpha_s)$ in perturbation theory. The six 
structure functions $F_1^{\nu,\bar\nu}$, 
$F_2^{\nu,\bar\nu}$ and $F_3^{\nu,\bar\nu}$ may in the parton 
model be expressed in terms of parton densities as
\bea
 F_1^\nu &=&\bar u + d + s + \bar c,
\quad\quad\quad\qquad F_1^{\bar\nu} = u + \bar d + \bar s + c,\nonumber\\
 F_2^\nu &=&2x(\bar u + d + s + \bar c),
\quad\qquad F_2^{\bar\nu} = 2x(u + \bar d + \bar s + c),\\
xF_3^\nu &=& 2x(-\bar u + d + s - \bar c),
\qquad xF_3^{\bar\nu} =2x(u -\bar d - \bar s + c),\nonumber
\label{part}
\eea
where we have set the CKM mixing angles to zero for simplicity, and 
restricted attention to the first four flavors.\footnote{It is 
inappropriate to consider only three flavors in charge current 
scattering since scattering off a strange quark produces a charmed 
quark in the final state.} To go from a proton to a neutron target
(assuming isospin invariance) we interchange $u$ and $d$. It is
thus not difficult to see that by constructing appropriate 
linear combinations of all eight independent structure functions
(conventionally taken as $(F_2^{\nu,\bar\nu})_{p,n}$ and 
$(xF_3^{\nu,\bar\nu})_{p,n}$)  
obtained by $\nu$ and $\bar\nu$ scattering on proton and neutron 
(or deuteron) targets it is possible to separately 
disentangle $u\pm \bar u$, $d\pm \bar d$ and 
$s\pm\bar s$ provided only that we can determine $c\pm\bar c)$,
either theoretically, or else empirically by tagging charm in neutral 
current processes (as is done currently at HERA). 
More explicitly, assuming isospin invariance between proton 
and neutron targets we have 
\bea
(F_2^{\nu+\bar\nu})_p=(F_2^{\nu+\bar\nu})_n
&=& x(u+\bar u + d+\bar d + s+\bar s + c+\bar c),\nonumber\\ 
(xF_3^{\nu-\bar\nu})_p-(xF_3^{\nu-\bar\nu})_n 
&=& -2x( u+\bar u-( d + \bar d)),\nonumber\\
(xF_3^{\nu-\bar\nu})_p+(xF_3^{\nu-\bar\nu})_n 
&=&2x( s+\bar s-( c + \bar c)),\nonumber\\
(xF_3^{\nu+\bar\nu})_p=(xF_3^{\nu+\bar\nu})_n
&=& x(u-\bar u + d-\bar d + s-\bar s + c-\bar c),\\ 
(F_2^{\nu-\bar\nu})_p-(F_2^{\nu-\bar\nu})_n 
&=& -2x( u-\bar u-( d - \bar d)),\nonumber\\
(F_2^{\nu-\bar\nu})_p+(F_2^{\nu-\bar\nu})_n 
&=&2x( s-\bar s-( c - \bar c))),\nonumber
\label{decomunpol}
\eea
where $F_i^{\nu\pm\bar\nu}\equiv \half (F_i^\nu \pm F_i^{\bar\nu})$.
The first and fourth of these equations are the structure functions 
$F_2^{\nu+\bar\nu}$ and $F_3^{\nu+\bar\nu}$ 
normally measured in neutrino scattering (though on heavy targets), 
while the second and third allow flavour decomposition of the total 
$q+\bar q$ distributions, and the fifth and sixth a similar decomposition 
for the valence distributions. To separate out 
strangeness from intrinsic charm empirically would require 
either a tagging of charm in the final state to give an independent 
determination of strangeness alone, or (in principle at least) a 
combined analysis with neutral current structure function data.

\begin{figure}[t!]
\hspace{3.0cm}
\psfig{figure=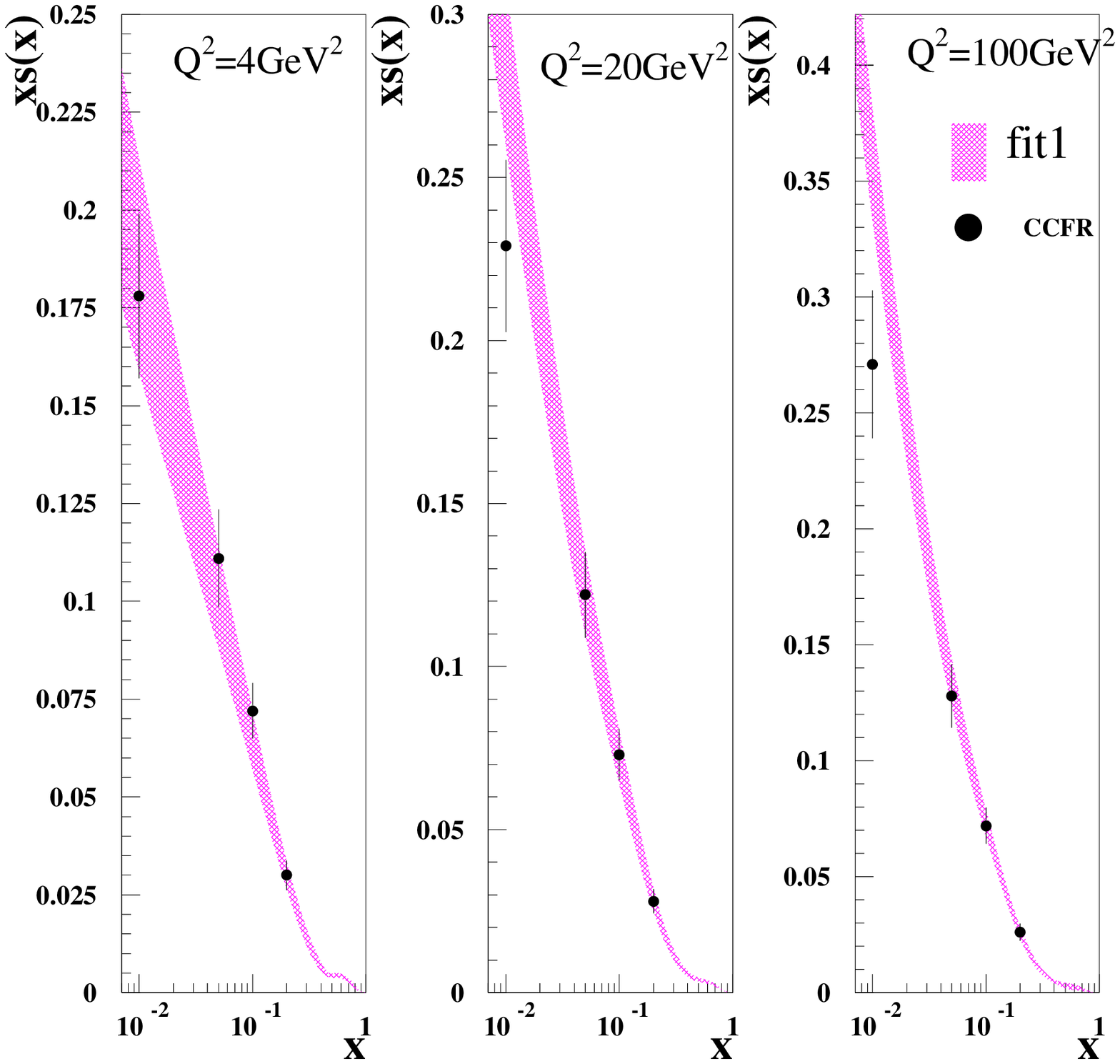,height=3.0in}
\caption{\label{fig:spsb}
A strange quark distribution extracted from charged current cross 
section data \cite{BPZ}: results from a NLO CCFR dimuon 
determination\cite{dimuon} are shown for comparison.
}
\end{figure}

In perturbative QCD it is possible to perform a complete NLO 
global analysis using well known results for NLO coefficient 
functions and anomalous dimensions. The charm contribution can 
then be computed perturbatively (on the assumption that the intrinsic 
charm is very small). Since only $F_2^{\nu+\bar\nu}$ contains 
a singlet component, while $F_2^{\nu-\bar\nu}$ and 
$F_3^{\nu\pm\bar\nu}$ are entirely nonsinglet, a 
clean extraction of both $\alpha_s$ \cite{alf} and the gluon 
distribution from scaling violations is 
possible with data taken over a sufficiently wide kinematic range.
To perform such an analysis it is necessary to either extract
the structure functions from the differential cross section in
a model independent way \cite{Bod}, or simply to fit directly 
to the cross section \cite{BPZ}. Fits to BEBC,CDHS and CDHSW data
suggest that it is already possible to extract $s+\bar s$ (see fig.
\ref{fig:spsb})
and possibly even $s-\bar s$: it will be interesting to see the 
results of this type of analysis applied to the much more 
precise CCFR/NuTeV data.

\section{Polarized Structure Functions}

Polarized structure functions may be defined in analogy with the 
unpolarized ones through asymmetries in the polarized 
cross-sections: for longitudinal polarization we may write
\bea
\smallfrac{d^2\Delta\sigma^{\nu,\bar\nu}}{dx dy}&=&
\smallfrac{G_F^2 S}{\pi(1+Q^2/M_W^2)^2}
[\pm y(1-\smallfrac{y}{2}-\smallfrac{xym}{2E})2xg_1^{\nu,\bar\nu}
\mp \smallfrac{2x^2ym}{E}g_2^{\nu,\bar\nu}\nonumber\\ 
&+&(1-y-\smallfrac{xym}{2E})(g_4^{\nu,\bar\nu}
+\smallfrac{xm}{E}(g_4^{\nu,\bar\nu}-g_3^{\nu,\bar\nu}))
+y^2x(1+\smallfrac{xm}{E})g_5^{\nu,\bar\nu}
],\label{psfn}
\eea
where the polarization asymmetry $\Delta\sigma
=\sigma^\leftarrow_\Rightarrow-\sigma^\leftarrow_\Leftarrow$
(the unpolarized cross-section (\ref{sfn}) being 
$\sigma= \half(\sigma^\leftarrow_\Rightarrow 
+\sigma^\leftarrow_\Leftarrow)$, which accounts for the extra 
factor of $2$ on the \rhs\ of (\ref{psfn})).
With these definitions\cite{bk}
\footnote{There are many variants 
in the literature: see \cite{bk} for a compilation.} 
in the high energy limit $E\gg m$ $g_2$ and $g_3$ drop out,
and we are left with an expression of the same form as the 
unpolarized decomposition (\ref{sfn}), but with $F_1\to g_5$, 
$F_2\to g_4$ and $F_3\to 2g_1$. Thus we again have six 
partonic structure functions for every target and for 
high energy polarization asymmetries  
the complicated decomposition (\ref{psfn}) then 
becomes simply
\beq
\smallfrac{d^2\Delta\sigma^{\nu,\bar\nu}}{dx dy}=
\smallfrac{G_F^2 S}{\pi(1+Q^2/M_W^2)^2}
[\pm y(1-\smallfrac{y}{2})2xg_1^{\nu,\bar\nu} 
+(1-y)g_4^{\nu,\bar\nu}+y^2xg_5^{\nu,\bar\nu}].\label{psfnhe}
\eeq

The remaining four structure functions $g_2^{\nu,\bar\nu}$ and 
$g_3^{\nu,\bar\nu}$ have no simple partonic interpretation
and are contaminated by twist three contributions: their twist 
two components are fixed by the Wandzura-Wilczek relation
(giving $g_2$ in terms of $g_1$) and a similar relation \cite{bk} 
which gives $g_3$ in terms of $g_4$. They are most easily 
determined by measuring transverse asymmetries: such measurements 
are very difficult however because the transverse asymmetry is 
suppressed by $m/Q$.

In the parton model $g_4$ and $g_5$ are related by an 
analogue\cite{Dicus} of the Callan-Gross relation: 
$g_4 = 2x g_5(1+O(\alpha_s))$.  
The flavor decomposition of the structure 
functions $g_1^{\nu,\bar\nu}$,  $g_4^{\nu,\bar\nu}$ and 
$g_5^{\nu,\bar\nu}$ may thus be expressed in terms of 
parton densities as
\bea
g_1^\nu &=&  \Delta\bar u + \Delta d + \Delta s 
+ \Delta\bar c,
\qquad\qquad\quad g_1^{\bar\nu} = \Delta u +\Delta\bar d + 
\Delta\bar s + \Delta c,\nonumber\\
g_4^\nu &=& 2x(-\Delta\bar u  + \Delta d +\Delta s - \Delta \bar c),
\qquad g_4^{\bar\nu} = 2x(\Delta u - \Delta \bar d 
- \Delta \bar s + \Delta c),\\
g_5^\nu &=& -\Delta\bar u  + \Delta d +\Delta s - \Delta \bar c,
\qquad\qquad g_5^{\bar\nu} = \Delta u - \Delta \bar d 
- \Delta \bar s + \Delta c,\nonumber
\label{polpart}
\eea
in precise analogy with the unpolarized case (\ref{part}): comparing 
(\ref{sfn}) with (\ref{psfnhe}), 
$F_1\rightarrow g_5$, $F_2\rightarrow g_4$, 
$\half F_3\rightarrow g_1$ and $q\rightarrow\Delta q$, 
$\bar q\rightarrow -\Delta\bar q$ (changing 
a quark to an antiquark also flips its helicity). 
Again, by constructing appropriate 
linear combinations of all eight independent structure 
functions (conventionally taken as $(g_1^{\nu,\bar\nu})_{p,n}$ and 
$(g_5^{\nu,\bar\nu})_{p,n}$) obtained by longitudinally polarized $\nu$ 
and $\bar\nu$ scattering on proton and neutron 
(or deuteron) targets it is possible 
to separately disentangle $\Delta u\pm \Delta\bar u$,
$\Delta d\pm \Delta\bar d$ and $(\Delta s\pm\Delta\bar s)$ just 
as in eq.(\ref{decomunpol}). 

Some combinations of the polarized structure functions are of particular 
interest. For example, writing 
$g_i^{\nu\pm\bar\nu}\equiv \half(g_i^\nu\pm g_i^{\bar\nu})$, the 
first moment of
\beq
2(g_1^{\nu+\bar\nu})_p=2(g_1^{\nu+\bar\nu})_n
=\Delta u+\Delta\bar u +\Delta d+\Delta\bar d 
+\Delta s+\Delta\bar s +\Delta c+\Delta\bar c
\label{decompola} 
\eeq
is the axial singlet charge $a_0$. This is a much more direct measurement 
than the traditional one through electron-proton or deuteron DIS 
since in the latter case one must first subtract the octet charge 
$a_8$ which is then only determined indirectly through hyperon decays 
(see \cite{bt} for a recent review). Thus in $\nu$-DIS one would 
have a direct check on the anomalous suppression of $a_0$. Similarly
first moments of
\bea
(g_5^{\nu-\bar\nu})_p-(g_5^{\nu-\bar\nu})_n 
&=& -(\Delta u+\Delta\bar u) +\Delta d + \Delta\bar d,\nonumber\\
(g_5^{\nu-\bar\nu})_p+(g_5^{\nu-\bar\nu})_n 
&=&\Delta s+\Delta\bar s-(\Delta c + \Delta\bar c),
\label{decompolb} 
\eea
give direct measurements of the axial charge $a_3$ (again currently only 
measured indirectly through $\beta$-decay) and of the contribution of strange 
quarks to the nucleon spin, as would the 
tagging of charm in the final state. 
Flipping the signs, we can also determine the contribution of 
valence quarks to the spin, since
\bea
2(g_5^{\nu+\bar\nu})_p=2(g_5^{\nu+\bar\nu})_n
&=&\Delta u-\Delta\bar u +\Delta d-\Delta\bar d 
+\Delta s-\Delta\bar s +\Delta c-\Delta\bar c,\nonumber\\ 
(g_1^{\nu-\bar\nu})_p-(g_1^{\nu-\bar\nu})_n 
&=& -(\Delta u-\Delta\bar u)+\Delta d - \Delta\bar d,\\
(g_1^{\nu-\bar\nu})_p+(g_1^{\nu-\bar\nu})_n 
&=&\Delta s-\Delta\bar s -(\Delta c - \Delta\bar c),\nonumber
\label{decompolc} 
\eea  
so one could even check for intrinsic strange polarization 
$\Delta s-\Delta\bar s$. None of these valence polarizations can be 
cleanly measured in current polarization experiments.

In practice this flavor separation would be best performed by a 
global fit in NLO perturbative QCD: all the NLO anomalous dimensions
\cite{NLO} and coefficient functions \cite{dfs,swv,ks} are known, the 
latter for heavy quarks, so the polarized charm contribution can 
be computed perturbatively. 

\begin{figure}[t!]
\psfig{figure=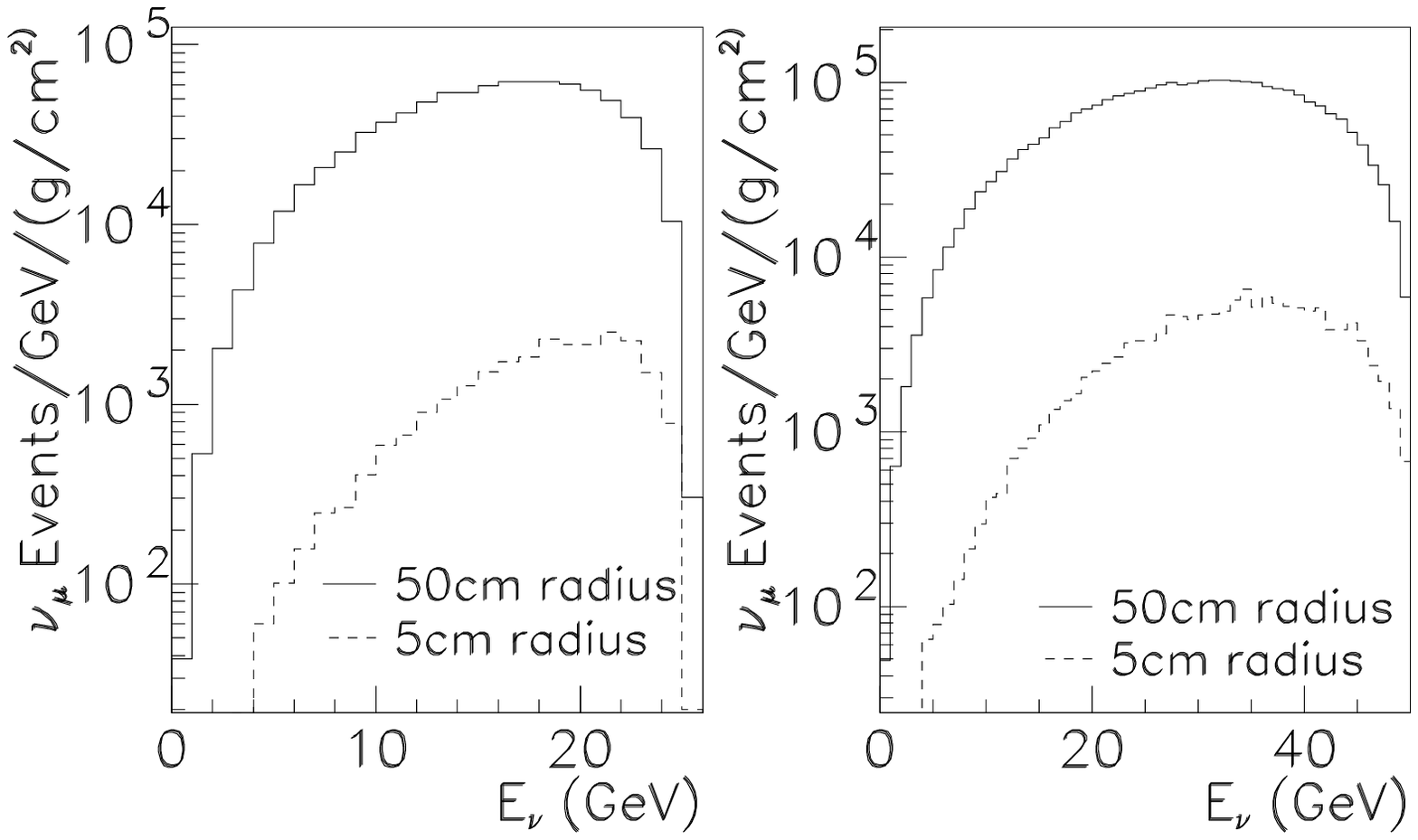,height=3.5in}
\caption{\label{fig:flux} Number of muon neutrino events per GeV per 
g/cm$^2$ 
for $10^{20}$ muon decays in an 800m long straight section followed 
by a 40m shielding section and a detector of radius 5 or 50 cm. Other  
storage ring parameters are given in the text.  } 
\end{figure} 

\section{Estimates of Neutrino interactions and Event Rates} 

     In order to estimate the precision with which the structure
functions above can be measured at a neutrino factory on nuclear targets, 
a GEANT-based Monte Carlo was used, with hydrogen and deuterium 
chosen as targets.  This simulation used LEPTO and JETSET 
(versions 6.5.1 and 7.408 respectively) to determine 
what the struck quark and its fragmentation were for 
each interaction, and the parton densities 
used were from CTEQ2MS \cite{cteq2ms}.  The hadron showers were 
not traced through the detector; because of the low densities of 
the targets and their sizes it is expected that most of the shower
will leave the target and be collected in a low mass particle-tracking 
system.  Therefore, in the results described below there is no smearing
from detector resolution.  The acceptance for muon neutrino and 
antineutrino charged current interactions is assumed to be 100\%  
for events with muons above 3~GeV, and to remove events from the 
quasielastic and resonance region only events with hadron energies 
above 1~GeV were considered.  It should be noted that the precisions 
listed here are pessimistic, since they are calculated assuming only an 
incoming $\nu_\mu$($\bar\nu_\mu$) flux.  The $\bar\nu_e$ 
($\nu_e$) fluxes that arrive simultaneously could in principle donate roughly 
another factor 
of two in event statistics.  However, since the acceptance and backgrounds
for $\nu_e$ charged current events are much more detector dependent 
these events are not considered here.  

     The question of where the high-rate neutrino physics experiments 
would occur at a neutrino factory is not a trivial one, but for 
a rough estimate of the overall statistics in neutrino ($\mu^-$ in the 
storage ring) and antineutrino ($\mu^+$ in the storage ring)  
running, consider two scenarios:  a 25 and 50~GeV muon storage
ring with 800m straight sections, followed by 30m of active shielding.  
Two target sizes were considered, namely a 5cm radius or a 50cm radius.  
In both cases the muon beam spot size was 
$1.23{\rm cm (x)}\times 0.883{\rm cm (y)}$, 
and the muon beam divergence was $0.73$~mrad in both the vertical and 
horizontal directions.  
Figure \ref{fig:flux} shows the muon neutrino fluxes for these 
different scenarios.  The (x,$Q^2$) regions accessible are comparable
between two different radii at the same energy: the most important factor 
is the loss of statistics for the smaller target.  
Note that the rate difference between the two radii is roughly 10 at 
high energies rather than a factor of 100.
Table~1 gives the neutrino charged current interaction 
rates for these different scenarios on a $1$~g/cm$^2$ isoscalar target.  

\begin{table}[!t] 
\label{tab:rates} 
\begin{center}
\begin{tabular}{|l|ll|}
\hline
& \multicolumn{2}{|c|}{Detector Radius} \\  
Muon Energy &  5cm & 50cm \\ 
\hline
25 GeV & 24.5K & 841K \\
50 GeV & 131K & 2900K \\ 
\hline
\end{tabular} 
\end{center}
\caption{Muon neutrino charged current interaction 
rates for $10^{20}$ muon decays (one year) for 
detectors of different radius and for storage 
rings of different energies.}  
\end{table} 

Given that the density of cryogenic
deuterium is at least 0.162~g/cm$^3$, 1~g/cm$^2$ corresponds 
to only $6$~cm of deuterium.  Targets 1.3 meters long of 
cryogenic (polarized!) material have 
already been used by SMC, so one can imagine multiplying the 
statistics listed in table~1 by a factor of 20 
and still have a 
10g/cm$^2$ target after fiducial cuts on the vertex.  Much larger targets 
of liquid Hydrogen have been used in the past 
in neutrino experiments, but with extremely low neutrino event 
statistics.  

The relative cross sections per nucleon between 
$\nu D_2 : \bar\nu D_2 : \nu H_2 : \bar\nu H_2 $ are 
approximately 2:1:1.3:1.3, (as calculated by GEANT)
due to the higher abundance of up quarks in $H_2$ compared to $D_2$.  
For an unpolarized target one can consider 
radii of 50cm.  Due to the strong B field and low temperature requirements 
for the polarized targets (as well as the small beam spot size), 
at present only polarized targets of 5cm diameter 
have been used for charged lepton scattering experiments \cite{smctarg}.  
Given the rate 
of advances in cryogenic and magnetic field technology, it is not
unreasonable to expect that much larger targets will be available
several years from now.  For that reason we  
consider targets of polarized materials that are the same size as 
the modest unpolarized targets.  

\begin{figure}[t!]
\psfig{figure=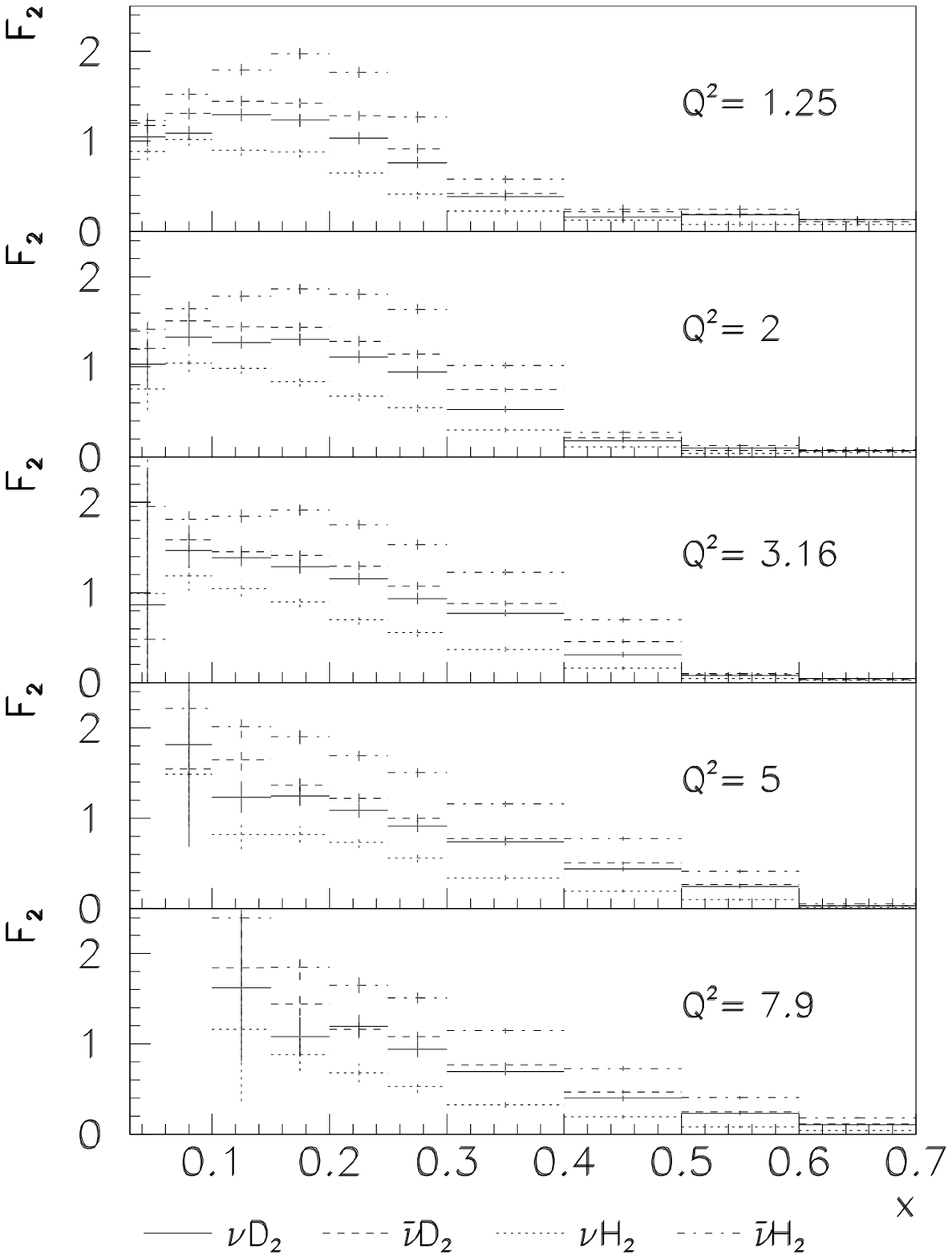,height=6.5in}
\caption{\label{fig:f2}
Expected results for $F_2^{\nu,\bar\nu}$ from one year of running in 
each mode, with a 0.1 g/cm$^2$ target: for a 10 g/cm$^2$ target 
the errors would be reduced by a factor of 10.} 
\end{figure} 
\begin{figure}[t!]
\psfig{figure=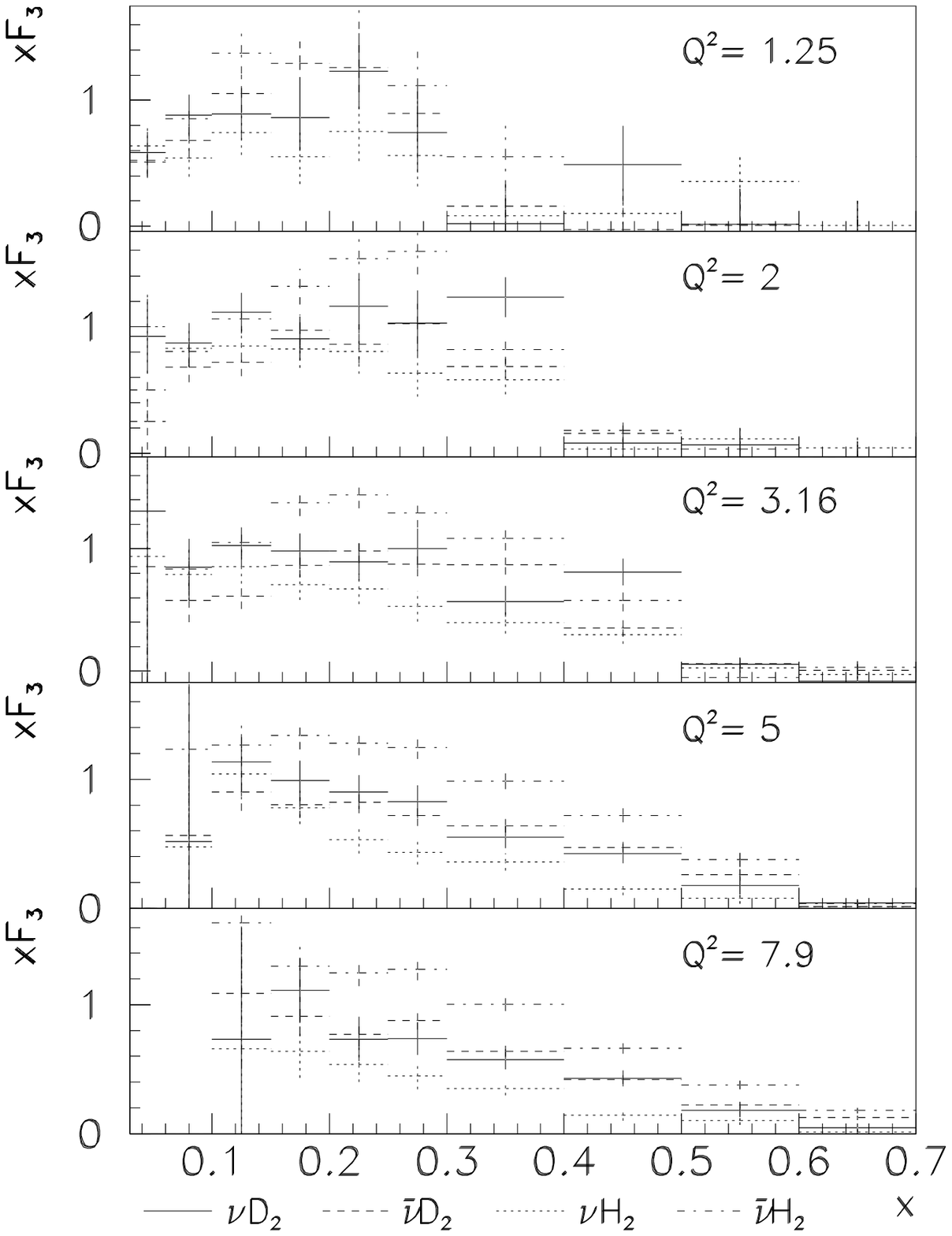,height=6.5in}
\caption{\label{fig:xf3}
The same as fig.\ref{fig:f2}, but this time showing results for $xF_3$.}
\end{figure} 
\begin{figure}[t!]
\psfig{figure=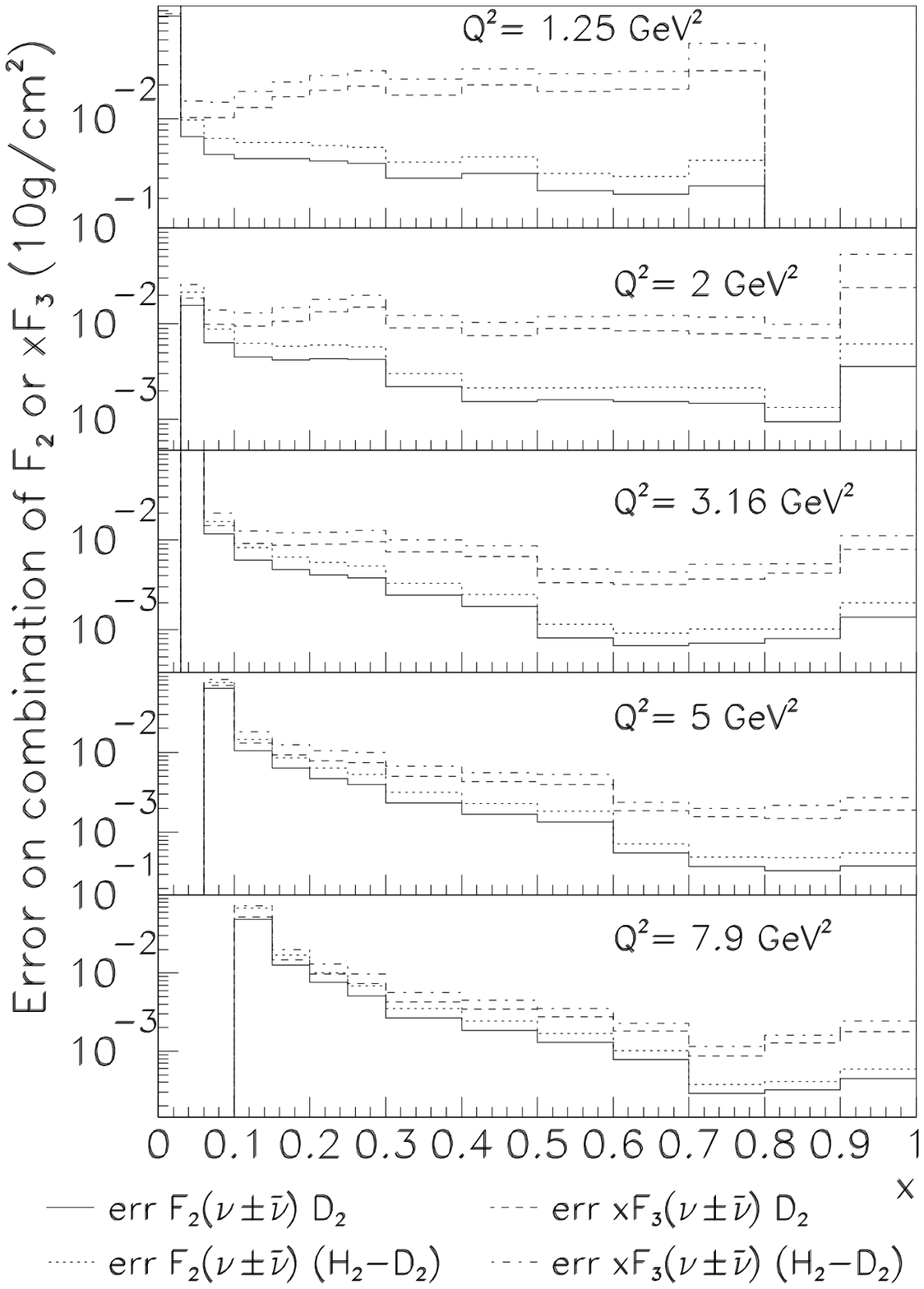,height=6.5in}
\caption{\label{fig:errs}
Estimated errors on $F_2^{\nu\pm\bar\nu}$ and $xF_3^{\nu\pm\bar\nu}$ 
after one year of running on each of 
neutrinos and antineutrinos, on a target of 10$g/cm^2$ 
of $D_2$ and $H_2-D_2$ for the 
range of x and $Q^2$ space above $Q^2=1~{\rm GeV}^2$.  
Note that one can achieve 
errors of better than 0.01 on sums and differences of 
structure functions with these statistics in most of the 
kinematic range accessed.  
}
\end{figure}

\section{Unpolarized Cross Sections at a Neutrino Factory}

To estimate the uncertainties on the unpolarized structure functions at 
a muon storage ring experiment, a leading order 
"model-independent" analysis was performed on 
the Monte Carlo data, as is described in \cite{Bod}.  
Simply put, the differential cross sections are measured in ($x,y,E$) 
bins, and then they are collected into ($x,y,Q^2$) bins, such that for a 
given ($x,Q^2$) bin there is a range of neutrino energies contributing
as $y$ goes from 0 to 1.  
The errors on the cross sections in each bin are taken to be the 
statistical errors alone resulting from the above fluxes 
(taking into account the total cross section differences between 
the different targets and probes).  For a given ($x,Q^2$) bin 
the $y$ distribution is fit to the function in eqn.(\ref{psfnhe}), 
setting $2xF_1=F_2$.  This will modify the central values fit since 
in reality the ratio $R=(F_2-2xF_1)/F_2$ is non-zero in 
neutrino scattering just as it is in electron scattering \cite{ukyangR}, 
but the errors on $F_2$ and $xF_3$ should not be affected.  
Also, there are no bin-centering corrections applied here, 
but the resulting fit errors should again not be affected.  

Figures \ref{fig:f2} and \ref{fig:xf3} show the fit results for the 
structure functions $F_2$ and $xF_3$ on both deuterium and hydrogen 
for both neutrinos and antineutrinos.  Because of limited Monte Carlo 
statistics, the errors shown above are for $0.1$ g/cm$^2$ targets,  
for $10^{20}$ 50 GeV muons decaying in the straight section in each 
mode ($\mu^+$ and $\mu^-$), and for 
a target of radius 50~cm.  The $\chi^2$ for the fits were roughly at 
$1$ per degree of freedom or better, because the Monte Carlo 
statistics were slightly  higher than the expected experimental 
statistics for these targets.  
The error reported by the fit does represent the experimental error 
expected, however, not the error due to the Monte Carlo statistics.  

These data access a $Q^2$ range that is lower than most of the 
CCFR data.  For the expected precision of a modest-sized target, one 
would divide these errors by a factor of 10 which assumes a 10 g/cm$^2$ 
fiducial target.  What makes these measurements unique however is 
the possibility 
of taking particular sums and differences between combinations of 
$\nu$ and $\bar\nu$, hydrogen and deuterium structure functions.  Figure 
\ref{fig:errs} shows the expected errors on the various
structure function combinations listed in eqn.\ref{decomunpol},
remembering that $D_2 = \half(p+n)$, while $(H_2-D_2) = \half(p-n)$.

The errors on the sums and differences of $F_2$ and $xF_3$ 
are comparable, but while the sums are of order unity, some of the 
differences are expected to be quite small.  Also, the 
errors on $xF_3$ are larger than those for $F_2$ for every 
combination, although they 
approach each other as $Q^2$ increases.  What is important here however 
is that the errors are uniformly small; less than $0.01$ for most 
of the ($x,Q^2$) range accessible.  For comparison, the statistical 
errors on $(\nu+\bar\nu)$ Fe structure functions from CCFR in its 
kinematic range vary from 0.01 to 0.04 \cite{alf}.

\section{Polarized Cross Sections at a Neutrino Factory} 

The polarized cross section measurements will be more 
difficult due to several factors, Firstly,  to keep the target polarized it
must be in both a strong magnetic field and a very low temperature
container, so the target size may be limited to smaller volumes 
than for unpolarized targets. Secondly, since one is taking 
differences of cross sections the polarized cross-sections 
will be smaller than the unpolarized ones and the fractional error 
correspondingly larger.  Finally, the fraction of 
polarized nuclei will necessarily be lower than 100\%.  
Two possible polarized target materials used in the past in 
charged lepton scattering are polarized 
solid butanol (used by the SMC collaboration) \cite{smctarg} 
or a polarized 
HD target (used by the LEGS collaboration) \cite{hdnim}.  

If one only polarizes the hydrogen or the deuterium in the target 
sample then when one takes the differences between opposite polarizations 
the result will only depend on the polarized nuclei. In this 
way one can use deuterated butanol and butanol to measure 
the D and the H cross section differences.  In the HD target, the two 
components are independently polarizable.  So for measurements of 
p+n or D alone, one can simply polarize the D in the sample and leave the H 
unpolarized.  For measurements of p-n, one can either polarize only 
the H, or use targets where 
the the H and the D are polarized in opposite directions. In the latter
case one is effectively scattering off polarized neutrons.  

To see how the errors on the polarized cross sections compare to 
those on the unpolarized cross sections, consider the following
argument:  for a perfectly polarized $H_2$ target of the same size 
as that considered above ($10$g/cm$^2$, 50cm radius) if all the protons
were polarized, then the error on the cross section difference 
would rougly be $1/\sqrt{2}$ times the error on the unpolarized case, 
assuming one integrated $10^{20}$ muon decays in one polarization and 
$10^{20}$ muon decays in the opposite polarization.  Factors such as 
the incomplete polarization of the target and the fact that the polarized
target is not 100\% $H_2$ enter into the error in the following way: 
\bea 
\sigma_{\rm pol} & = & \frac{\sigma_{\rm unpol}}{f_{\nu ,\bar\nu}P} 
\sqrt{\frac{\rho_{\rm unpol}}{\rho_{\rm pol}}}
\eea  
where $\rho_{\rm unpol}/\rho_{\rm pol}$ is the ratio of target 
densities to $H_2$ or $D_2$, $f_{\nu,\bar\nu}$ is the dilution 
factor and $P$ is the polarization of the $H_2$ or $D_2$.
The dilution factor $f_{\nu ,\bar\nu}$ is the $\nu$ (or $\bar\nu$) 
cross-section weighted ratio of the 
polarized nucleon to total nucleon content of the target.  
Because the different targets have of necessity different 
ratios of protons and neutrons, the neutrino and antineutrino 
dilution factors will be different.  

Table 2 compares the SMC and LEGS targets for both 
the $D_2$ cross-section difference and the $H_2$ cross-section difference 
\cite{smctarg}\cite{hdnim}\cite{honig}. 
In summary, in the SMC target the density is 
high but the dilution factor is small, while for the HD target the 
density is low but the dilution factor is high.  The HD target is newer
and has not been made in as large samples as the SMC target, but the
HD targets are modular and one can easily imagine adding 
several together.  
Overall the multiplicative factors coming from target details alone 
range from $1.6$ to $4.8$.  

\begin{table}
\label{tab:poltarg}
\begin{center}
\begin{tabular}{|l|l|l|l|l|}
\hline 
Characteristic & p-Butanol & D-butanol & HD(D$\Uparrow$) & HD (H$\Uparrow$D$\Downarrow$)\\
\hline
Density ($g/cm^3$) & 0.61 & 0.69 & 0.05 & 0.05 \\ 
D/H Polarization &  86\% & 51\% & 70\% & 70\%(D), \\ 
                                &  & & & 95\%(H) \\ 
Dilution Factor $\nu$ 	& 0.15  & 0.22 &  0.4  & 1.0 \\
Dilution Factor $\bar\nu$ & 0.25 & 0.22 & 0.55   & 1.0 \\ 
Length (cm) & 120 & 120 & 10 & 10 \\ 
($g/cm^2$) & 73.2 & 82.8 & 0.5 & 0.5 \\ 
Diameter (cm) & 5 & 5 & 3 & 3 \\ 
B Field & 2.5 T & 2.5T & 7T & 7T \\ 
Temperature & 0.1K & 0.1K & 1.5K & 1.5K \\ 
total target factor ($\nu$) & 2.6  & 4.4  & 4.3 &  2.6 \\
total target factor ($\bar\nu$) & 1.6  & 4.4  & 3.1  & 4.8 \\ 
\hline
\end{tabular} 
\end{center}
\caption{Comparison of four polarized targets: p-butanol, D-butanol, 
an HD target with only the D polarized, or HD with both H and D polarized 
oppositely. The `density' is the effective density for solid butanol 
(which depends on the packing fraction), and the dilution factors 
were calculated based on neutrino and antineturino cross sections on 
protons and isoscalar nuclei. The total target factor only takes into 
account the ratio between the given material and liquid $H_2$ or $D_2$
densities, the polarization, and the dilution factor.} 
\end{table} 

\begin{figure}[t!]
\psfig{figure=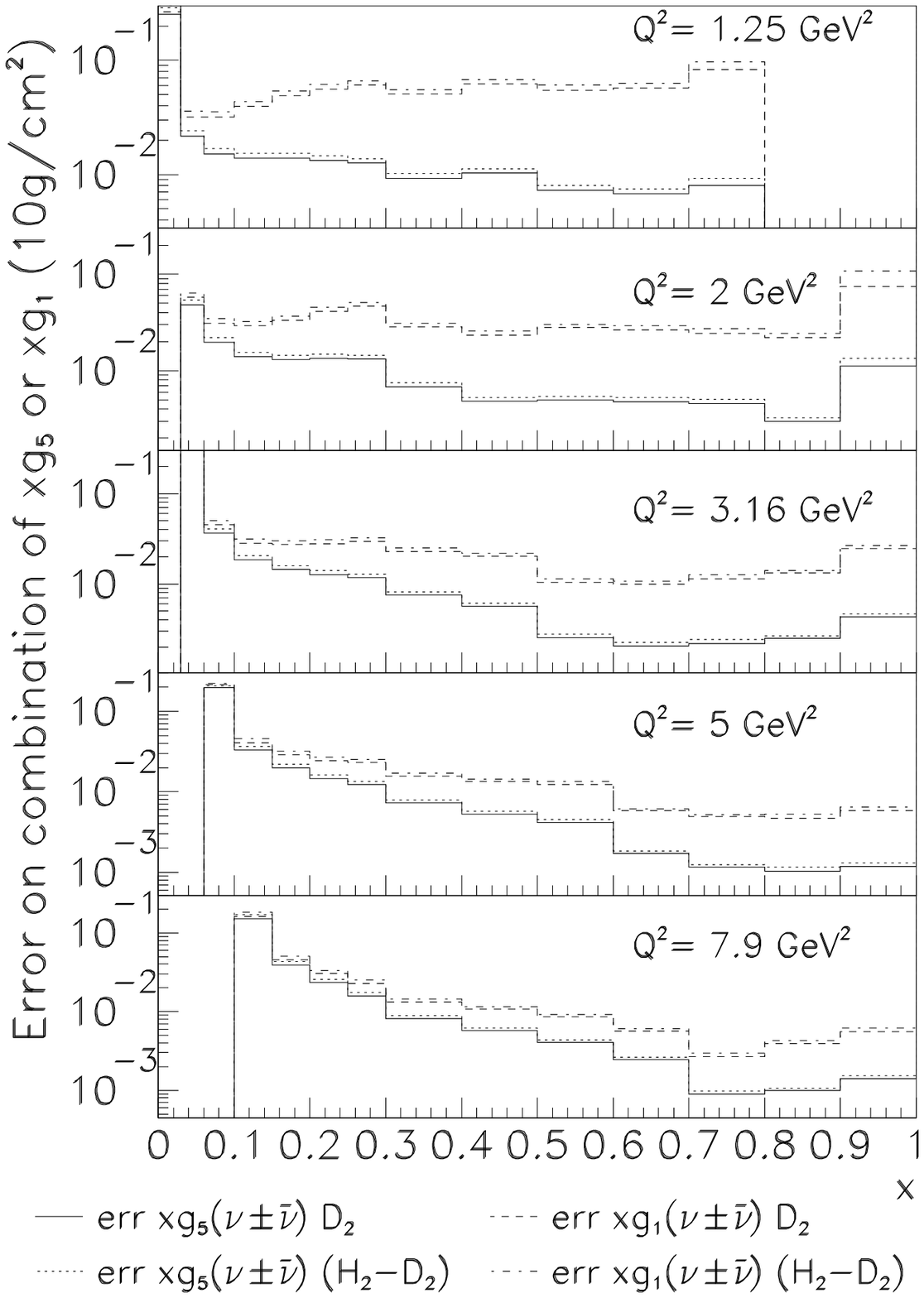,height=6.5in}
\caption{\label{fig:polerrs}
Error on polarized structure functions  on $xg_1^{\nu\pm\bar\nu}$ and 
$xg_5^{\nu\pm\bar\nu}$ coming from one year of running for each 
combination of neutrinos, antineutrinos,  polarized solid 
butanol and deuterated butanol in 2 polarizations for one 
small range of x and $Q^2$ space.  Note that one can achieve 
0.05 errors or better on spin structure functions with these statistics.  
}
\end{figure}

To translate between the cross section uncertainty and the 
polarized structure function uncertainty, recall that the 
functional form (up to terms in $m/E$ which are going to be small 
even for a $50$ GeV muon storage ring) for the unpolarized cross 
sections (eqn. \ref{sfn}) is the same as that for the polarized 
cross sections (eqn. \ref{psfn}), 
up to the substitutions $F_2 \to g_4=2xg_5$ and $F_3 \to 2g_1$.  
Therefore, the error on $xg_1$ or $xg_5$  for a given 
$x,Q^2$ bin is the same as that of $xF_3/2$ or $F_2/2$ 
in the unpolarized case multiplied by the ratio 
of the errors in the cross sections and the extra $\sqrt{2}$ from 
having two targets, described above.  

Figure \ref{fig:polerrs} shows the resulting 
errors on the various linear combinations of polarized structure
functions,  listed in eqn.\ref{decomunpol}, 
remembering again that $D_2 =\half(p+n)$, while $H_2-D_2 =\half(p-n)$. 
We have assumed one year's running at each polarization, each 
neutrino flavor, and each target.  Note that at the very least 
it is assumed that one would run in two target polarizations 
simultaneously, so this represents a total of 4 to 8 
years of running at $10^{20} \mu^\pm$ decays per year.  The 
targets are the same size as in figure \ref{fig:errs} but filled with  
either butanol or deuterated butanol, as was done by SMC \cite{smctarg}. 

The errors on the sums and differences of $xg_1$ and $xg_5$ 
are comparable, but while the sums are of order unity, some of the 
differences are again expected to be quite small.  Also, the 
errors on $xg_5$ are smaller than those for $xg_1$ for every 
combination, although they approach each other as $Q^2$ increases.  
What is important is that the errors are uniformly small, just 
as in the unpolarized case: less than a few times $0.01$ for most 
of the accessible ($x,Q^2$) range. 

\section{Conclusions}

We have examined the novel structure function measurements 
that could be made at a neutrino factory, and provided estimates
of their precision as a function of $x$ and $Q^2$ for nominal 
storage ring running ($10^{20}$ muons per year in a 50~GeV muon storage
ring). Because of the clean separation between valence and sea
afforded by neutrino and antineutrino running and the possibility 
of using both deuterium and hydrogen targets, such experiments 
could at last determine flavor by flavor the valence and sea quark 
distribution functions with statistical errors uniformly of order 
0.01 in each bin.  Systematic errors due to nuclear effects and 
beam energy would be minimal. Furthermore, by running with 
polarized targets already developed by the charged lepton scattering 
community the spin components of the proton, quark by quark, could be 
determined for the first time, again with statistical errors uniformly 
of order a few times 0.01. Such measurements would resolve definitively
the questions raised by the EMC experiment, in particular by 
determining both the strange and gluon contributions simultaneously 
in one experiment. Combining this data with complementary data from
eRHIC or polarized HERA could result in precise tests of our understanding 
of the spin structure of the nucleon.

A 50~GeV neutrino factory would be an ideal partonometer, revealing the 
partonic structure of the nucleon in exquisite detail. We hope that it
may be possible to enjoy such a facility in the not too distant future.

{\bf Acknowledgement:}\\  
RDB would like to thank M.~Mangano for discussions and correspondence 
on this subject, and S.~Forte for comments on the completed manuscript. 
DAH would like to thank M.~Velasco for discussions and help on this 
subject, in particular concerning spin physics and novel target 
techniques. This work was
supported in part by EU TMR  contract FMRX-CT98-0194 (DG 12 - MIHT). 

\end{document}